\documentclass[aps,prb,showpacs,twocolumn]{revtex4-1}
\usepackage{amssymb}
\usepackage{amsmath}
\usepackage{graphicx}
\usepackage{epsfig}
\usepackage{subfigure}
\usepackage{appendix}

\begin{document}

\title{Quantum transport in non-Hermitian impurity array}
\author{K. L. Zhang}
\author{X. M. Yang}
\author{Z. Song}
\email{songtc@nankai.edu.cn}
\affiliation{School of Physics, Nankai University, Tianjin 300071, China}
\begin{abstract}
We study the formation of band gap bound states induced by a non-Hermitian
impurity embedded in a Hermitian system. We show that a pair of bound states
emerges inside the band gap when a parity-time ($\mathcal{PT}$) imaginary
potential is added in a strongly coupled bilayer lattices and the bound
states become strongly localized when the system approaches to the
exceptional point (EP). As a direct consequence of such $\mathcal{PT}$
impurity-induced bound states, an impurity array can be constructed and
protected by energy gap. The effective Hamiltonian of the impurity array is
non-Hermitian Su-Schrieffer-Heeger (SSH) type and hosts Dirac
probability-preserving dynamics. We demonstrate the conclusion by numerical
simulations for the quantum transport of wave packet in right-angle bends
waveguide and $Y$-beam splitter. Our finding provides alternative way to
fabricate quantum device by non-Hermitian impurity.
\end{abstract}

\maketitle

\section{Introduction}

\label{Introduction} Throughout physics, stable or equilibrium phenomena can
be understood by bound state, ranging from quantum to classical objects. In
quantum physics, a bound state is a localized state of a particle subject to
a real valued potential, which may be the result of the presence of another
particles. The concept of bound state is ubiquitous in numerous branches of
physics, such as optics and condensed matter. In experiment, engineering
bound state can be generated by artificial defects in photonic crystals and
an array of defects, which are known as coupled-resonator optical waveguides
providing almost lossless guiding, bending of wave packet\cite%
{Joannopoulos,John,Yablonovitch,Skorobogatiy,Engelen,Hughes,Kuramochi,Thomas,Mazoyerprl,Mazoyeroe,Faolain}%
. By introducing artificial defects, various photonic crystal devices can be
realized for the applications in a wide variety of fields. Nowadays a
complex potential is not forbidden since non-Hermitian quantum mechanics has
emerged as a versatile platform for fabricating functional devices in
non-Hermitian regime. The main mechanism is based on the existence of
imaginary potential which has been investigated theoretically \cite%
{Bender,JPA2,Ali,Znojil,Jones,OL07,PRL08a,PRL08b,Joglekar10,Joglekar11,YDChong,HJingPRL2014}
and realized in experiment \cite%
{AGuo,CERuter,Wan,Sun,LFeng,BPeng,LChang,LFengScience,HodaeiScience,NC2015}
as an ideal building block of non-Hermitian systems. A non-Hermitian term
has a distinguishing feature that differs from a Hermitian one can be
exemplified by a simple two level system. An extra pseudo-Hermitian
non-Hermitian term always shrinks the level spacing, while a nontrivial
Hermitian perturbation always leads to the repulsion of the two levels.
Presumably, mid-gap levels can be generated by adding non-Hermitian impurity
on a Hermitian gapped system.

In this paper, we study the possibility of a quantum channel generated by an
array of non-Hermitian defects. Based on the Bethe ansatz solution we show
that a pair of bound states emerges inside the band gap when a parity-time ($%
\mathcal{PT}$) imaginary potential is added in a strongly coupled bilayer.
Such bound states are protected by energy gap, especially as the bound-state
energy tends to zero when the system approaches to the exceptional point
(EP). Consequently, an impurity array can be constructed in the mid-gap and
provides a low-loss waveguide since other Hermitian degrees of freedom are
adiabatically eliminated when we consider the dynamics in the impurity
array. In addition, we show that the effective Hamiltonian of the impurity
array is equivalent to a non-Hermitian Su-Schrieffer-Heeger (SSH) system and
obeys chiral-time ($\mathcal{CT}$) symmetry. This ensures quasi
orthogonality of the mid-gap modes in the framework of Dirac inner product,
and therefore the Dirac probability-preserving dynamics in the waveguide.
Therefore, although the waveguide is engineered by non-Hermitian impurities,
it acts as a conditional Hermitian device. We demonstrate the conclusion by
numerical simulations for the quantum transport of wave packet in
right-angle bends waveguide and $Y$-beam splitter. Our finding provides
alternative way to fabricate quantum device by non-Hermitian impurity.

This paper is organized as follows. In Sec. \ref{Formalism}, we present the
main ideas of non-Hermitian-impurity induced waveguide. In Sec. \ref{Bilayer
square lattice}, we provide a concrete example to illustrate our theory.
Sec. \ref{Dynamics in waveguide} demonstrates the dynamics of waveguide in
the concrete system. Finally, our conclusion is given in Sec. \ref%
{Conclusion and Discussion}.

\section{Formalism}

\label{Formalism} 
\begin{figure}[h]
\centering
\includegraphics[width=0.48\textwidth]{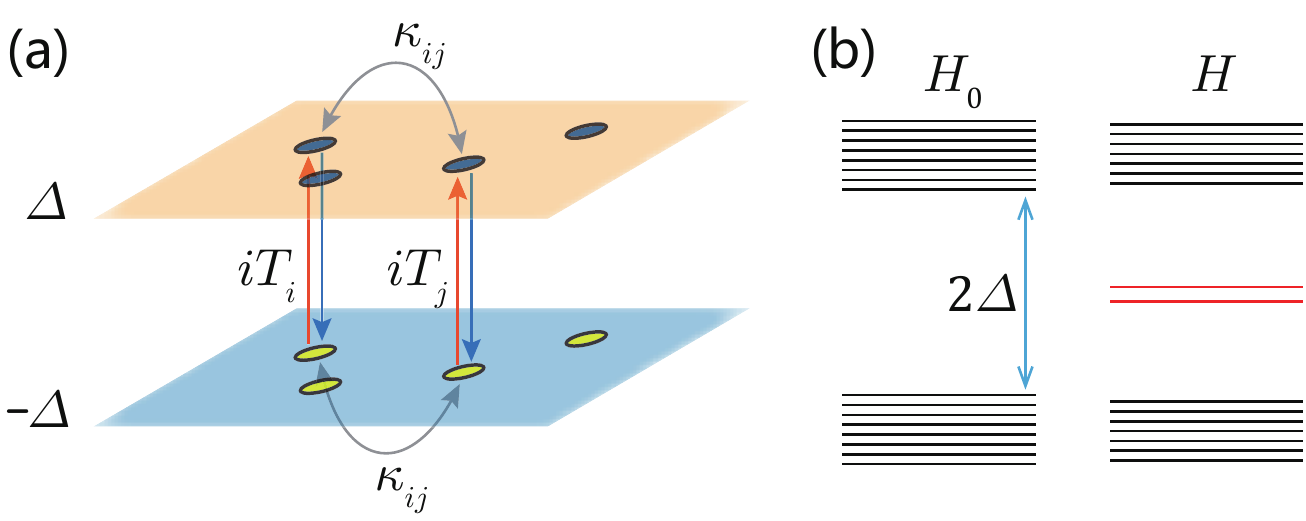}
\caption{(color online) (a) Schematic illustration of a two-layer
tight-binding lattice with non-Hermitian imaginary tunneling. (b) Band
structures for $H_{0}$ and $H$. In the absence of the interlayer tunneling,
a gap in order of $2\Delta $ opens. When a single tunneling $iT_{j}$\
switches on, two isolated energy levels can be achieved around the midgap.}
\label{Fig1}
\end{figure}

Consider a bilayer system (Fig. \ref{Fig1}(a)), composed
of two identical lattice but with opposite on-site energies. The interlayer
tunneling is non-Hermitian and sparse, which is referred to as non-Hermitian
impurity. The Hamiltonian is given by%
\begin{eqnarray}
H &=&H_{0}+H_{T},  \notag \\
H_{0} &=&\sum_{\left\langle i,j\right\rangle }\sum_{\sigma =\pm }\left(
\kappa _{ij}a_{i,\sigma }^{\dag }a_{j,\sigma }+\mathrm{H.c.}\right)
+\sum_{j}\sum_{\sigma =\pm }\sigma \Delta a_{j,\sigma }^{\dag }a_{j,\sigma },
\notag \\
H_{T} &=&i\sum_{j}T_{j}\left( a_{j,+}^{\dag }a_{j,-}+\mathrm{H.c.}\right) ,
\label{formalism_H}
\end{eqnarray}%
where $\sigma $\ corresponds to the $+$ and $-$\ layers, and $\left\{
iT_{j}\right\} $ is a set of imaginary intralayer hopping parameters. Fig. %
\ref{Fig1}(a) is the schematic illustration of the model. We consider the
region with $\Delta \gg \left\vert \kappa _{ij}\right\vert $, which ensures
the energy gap of order $\Delta $\ between two ($+$/$-$) enargy bands for
the system $H_{0}$. In the following, we show that $H_{T}$\ may induce local
states within the energy gap.

In the absence of $H_{T}$, the states of $H_{0}$\ are extended states
spreading the probability over the two layers. In the limit case with \ $%
\left\vert T_{j}\right\vert \sim \Delta \gg \left\vert \kappa
_{ij}\right\vert $, each non-Hermitian tunneling $iT_{j}$ may induce two
isolated energy levels around the midgap, which have the form $\pm \sqrt{%
\Delta ^{2}-\left( T_{j}\right) ^{2}}$. It is responsible for the
non-Hermiticity of the impurity, since a real tunneling cannot form the
isolated levels within the energy gap, but beyond the two bands. Fig. \ref%
{Fig1}(b) is the band structures for $H_{0}$ and $H$. We note that the
eigenvalues of $H$\ are real or imaginary, i.e., the EPs only appear at zero
energy. Considering the case with a single nonzero $T_{j}$, there are two
isolated levels within the gap and such two bound states coalesce to a
single state $(1/\sqrt{2})\left( a_{j,-}^{\dag }+ia_{j,+}^{\dag }\right)
\left\vert \mathrm{vac}\right\rangle $ at the EP when $\Delta =\left\vert
T_{j}\right\vert $. If there are many such nonzero tunneling $T_{j}$ within
the unbroken region and the corresponding bound states overlap with each
other, sub-bands\ within\ the gap will form and a non-Hermitian-impurity
induced waveguide is achieved for a array of impurity.

\begin{figure}[tbh]
\centering
\includegraphics[width=0.48\textwidth]{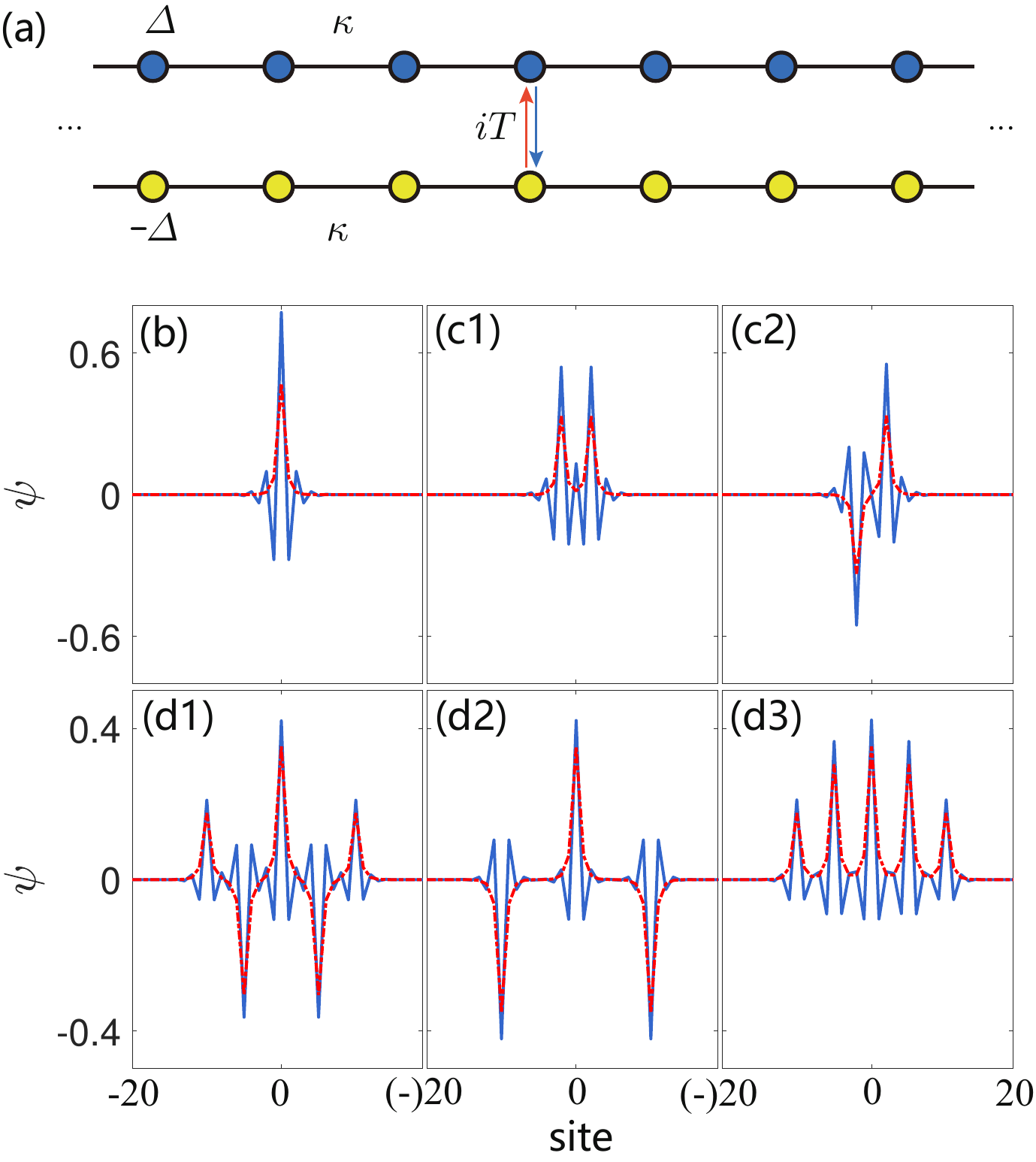}
\caption{(color online) (a) Schematic illustration of a bichain lattice with
opposite chemical potentials $\pm \Delta $ and a single non-Hermitian
tunneling $iT$. The intra- and interlayer hopping strengths are $\protect%
\kappa $ and $iT$, respectively. (b) Profile of one of the two bound states
for single imaginary tunneling $iT$. The ordinates of blue (solid) and red
(dash) lines are real and imaginary numbers, which represent the wave
function of the top and bottom chains, respectively. The parameters are $%
\Delta =5$, $\protect\kappa =1$ and $T=4$. (c1) and (c2) are profiles of two
of the four bound states for double imaginary tunneling $iT$. The parameters
are $\Delta =5$, $\protect\kappa =1$ and $T=4$. (d1)-(d3) are profiles of
three of the ten bound states induced by multi(five) impurities with
parameters $\Delta =5$, $\protect\kappa =1$ and $T=4$. The results are
obtained by numerical diagonalization and the wave functions are Dirac
normalized.}
\label{Fig2}
\end{figure}

To demonstrate the main idea, we consider an example model, which is consist
of two infinite chains with a single non-Hermitian tunneling. The
corresponding Hamiltonian has the form 
\begin{eqnarray}
H_{\mathrm{Chain}} &=&\sum_{j=-\infty }^{\infty }\sum_{\sigma =\pm }\left[
\left( \kappa a_{j,\sigma }^{\dag }a_{j+1,\sigma }+\mathrm{H.c.}\right)
+\sigma \Delta a_{j,\sigma }^{\dag }a_{j,\sigma }\right]  \notag \\
&&+iT\left( a_{0,+}^{\dag }a_{0,-}+a_{0,-}^{\dag }a_{0,+}\right) ,
\label{Chain_H}
\end{eqnarray}%
where $\sigma =+$ or $-$ is the index that respectively labels the position
in the top or bottom chains, and $j$ is the in-chain site index. Parameters $%
\kappa $ and $T$ of this model are intra- and interchain hopping strengths.
The schematic illustration is shown in Fig. \ref{Fig2}(a). Bethe ansatz
method shows that there are two bound states around the center with the
eigenvector (see Appendix 1) 
\begin{eqnarray}
\left\vert \psi _{\mathrm{B}}\right\rangle &=&C\sum_{j=-\infty }^{\infty
}\left( e^{i\pi j}e^{-\beta ^{+}\left\vert j\right\vert }a_{j,+}^{\dag
}\right) \left\vert \mathrm{vac}\right\rangle  \notag \\
&&-C\frac{2\kappa }{iT}\sinh \beta ^{+}\sum_{j=-\infty }^{\infty }\left(
e^{-\beta ^{-}\left\vert j\right\vert }a_{j,-}^{\dag }\right) \left\vert 
\mathrm{vac}\right\rangle .
\end{eqnarray}%
Here $C$ is the normalization coefficient, which is determined in the
context of Dirac or biorthogonal inner products. Profile of the bound state
is shown in Fig. \ref{Fig2}(b). The corresponding eigenenergy is%
\begin{equation}
E_{\mathrm{B}}=-2\kappa \cosh \beta ^{+}+\Delta =2\kappa \cosh \beta
^{-}-\Delta .
\end{equation}%
where $\beta ^{\pm }$ are positive real numbers and fulfill the equations 
\begin{equation}
\left\{ 
\begin{array}{c}
\cosh \beta ^{-}+\cosh \beta ^{+}=\Delta /\kappa \\ 
\sinh \beta ^{+}\sinh \beta ^{-}=T^{2}/(2\kappa )^{2}%
\end{array}%
\right. .
\end{equation}%
We note that two bound states coalesce to a single one 
\begin{equation}
\left\vert \psi _{\mathrm{B}}^{\mathrm{EP}}\right\rangle =C\sum_{j=-\infty
}^{\infty }\left( e^{i\pi j}e^{-\beta \left\vert j\right\vert }a_{j,+}^{\dag
}+ie^{-\beta \left\vert j\right\vert }a_{j,-}^{\dag }\right) \left\vert 
\mathrm{vac}\right\rangle ,
\end{equation}
at EP with $\beta =\ln \left[ \Delta /\left( 2\kappa \right) +\sqrt{\Delta
^{2}/\left( 4\kappa ^{2}\right) -1}\right] $ and $T=T_{\mathrm{c}}=\sqrt{%
\Delta ^{2}-(2\kappa )^{2}}$. When $T>T_{\mathrm{c}}$, the bound-state
energy becomes an imaginary number. This exact solution indicates that
stable bound states can be formed by the non-Hermitian impurity. In the case
of multi impurities, the above exact solution is still applicable when the
distance between two neighboring impurities is sufficiently large.
Nevertheless, the overlap of the wave functions occur and extended states
form within the region of the impurity array. Fig. \ref{Fig2}(c) and (d) are
plots of several typical bound states for two and five imaginary impurities.
For $n$ impurities, the system supports $2n$ bound states confined the
region of the array. Consequently, the mid-gap sub-bands may form for large $%
n$.

This result has many implications. (i) Similar to the Hermitian regime,
imaginary impurities can induce local bound states with real energies. (ii)
In contrast to a Hermitian impurity, energy levels of imaginary impurity can
be protected by band gap and coalesce at EP. (iii) Deliberately designed
impurity array can take the role of waveguide. It allows a variety of
non-Hermitian models with various of geometries and wide range of parameters
to be candidates of waveguides. In the next section, we will show its
application in an example.

\section{Bilayer square lattice}

\label{Bilayer square lattice} 
\begin{figure}[tbp]
\centering
\includegraphics[width=0.48\textwidth]{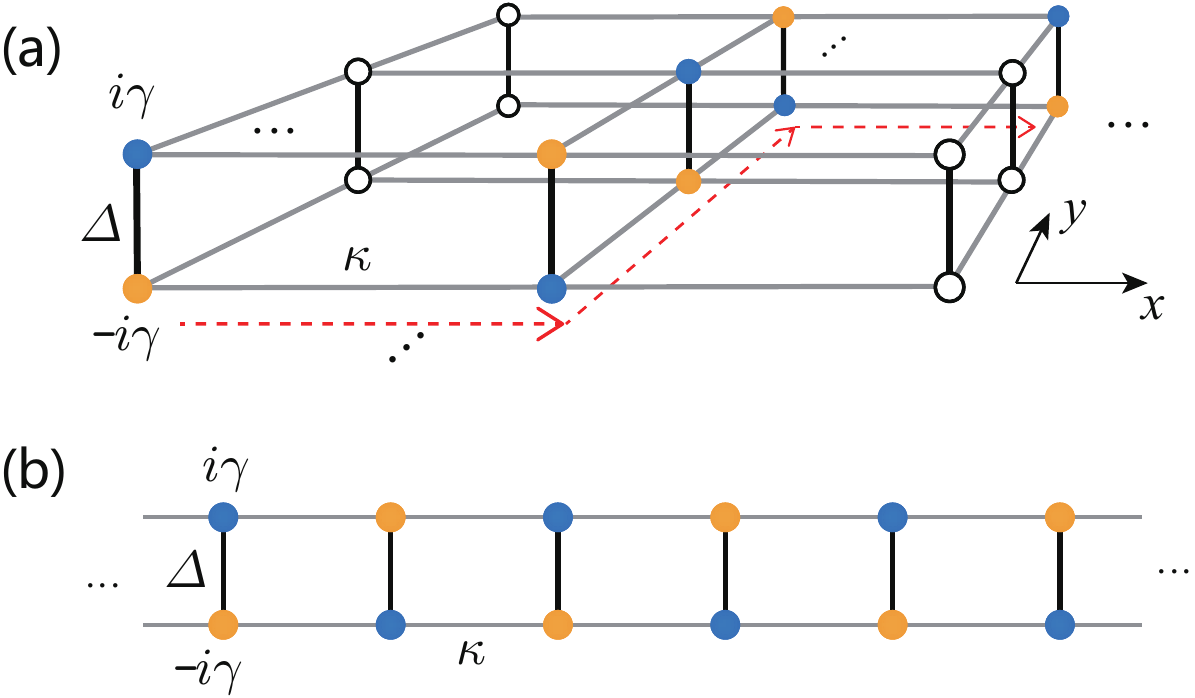}
\caption{(color online) (a) Schematic illustration of a bilayer square
lattice model with $\protect\kappa $ (gray line) and $\Delta$ (black line)
are intra and interlayer hopping strengths; the non-Hermitian imaginary
potentials $\pm i\protect\gamma$ are indicated by the blue ($i\protect\gamma$%
) and yellow ($-i\protect\gamma$) dots, and the white dots indicate zero
onsite potentials. (b) The waveguide path (red dash lines in (a)) forms a
non-Hermitian ladder model. }
\label{Fig3}
\end{figure}

The bound states induced by non-Hermitian impurities can be employed to
construct sub-bands in the energy gap, which constitute the channel for
specific transport of particle, as a waveguide in a Hermitian systems. In
the Hermitian realm, the waveguides of most kinds of discrete systems have
been\ well studied. It is usually be done by distroying the translational
symmetry. In the present work, our strategy is doing the same thing but by
adding non-Hermitian terms. In the following, we will present an example
that implements a 1D waveguide over a square lattice along any desired path.

We consider a bilayer square lattice model which is shown in Fig. \ref{Fig3}%
(a). The Hamiltonian can be written as the Hermitian part $H_{0}$ and
non-Hermitian part $H_{T}$,

\begin{equation}
H=H_{0}+H_{T}.
\end{equation}%
The corresponding Hermitian Hamiltonian has the form 
\begin{eqnarray}
H_{0} &=&H_{1}+H_{2}+H_{12},  \notag \\
H_{\lambda } &=&\kappa \sum_{j,l}\alpha _{j,l,\lambda }^{\dag }\left( \alpha
_{j+1,l,\lambda }+\alpha _{j,l+1,\lambda }\right) +\mathrm{H.c.},  \notag \\
H_{12} &=&\Delta \sum_{j,l}\alpha _{j,l,1}^{\dag }\alpha _{j,l,2}+\mathrm{%
H.c.},  \label{bilayer}
\end{eqnarray}%
where $\lambda =1$ or $2$ is the index that respectively labels the position
in the top or bottom layers, and $(j,l)$ is the in-plane site index.
Parameters $\kappa $ and $\Delta $ of this model are intra and interlayer
hopping strengths. And the distribution of imaginary potentials is given as
the form%
\begin{equation}
H_{T}=i\sum_{\lambda =1}^{2}\sum_{j,l}(-1)^{\lambda +j+l}\gamma _{jl}\alpha
_{j,l,\lambda }^{\dag }\alpha _{j,l,\lambda }.
\end{equation}%
By taking the linear transformations 
\begin{equation}
a_{j,l,\pm }^{\dag }=\frac{1}{\sqrt{2}}\left( \alpha _{j,l,1}^{\dag }\pm
\alpha _{j,l,2}^{\dag }\right) ,
\end{equation}%
the Hamiltonian Eq. (\ref{bilayer}) can be written as 
\begin{eqnarray}
H_{0} &=&\kappa \sum_{j,l}\sum_{\sigma =\pm }a_{j,l,\sigma }^{\dag
}(a_{j+1,l,\sigma }+a_{j,l+1,\sigma })+\mathrm{H.c.}  \notag \\
&&+\Delta \sum_{j,l}\sum_{\sigma =\pm }\sigma a_{j,l,\sigma }^{\dag
}a_{j,l,\sigma }.
\end{eqnarray}%
We note that the bond (antibond) state of a rung can only be transited to
the bond (antibond) state next to it with hopping strength $\kappa $.
Therefore it can be decomposed into two independent single layer square
lattices with on-site potentials $\Delta $\ and $-\Delta $, respectively.

Accordingly, the non-Hermitian term reads 
\begin{equation}
H_{T}=i\sum_{j,l}(-1)^{j+l+1}\gamma _{j,l}\left( a_{j,l,+}^{\dag }a_{j,l,-}+%
\mathrm{H.c.}\right) ,
\end{equation}%
which takes the role of interlayer imaginary tunneling. Obviously, the
present model is a concrete example of the system depicted in Eq. (\ref%
{formalism_H}). The obtained result is applicable to a wide kind of systems.

\begin{figure*}[tbh]
\centering
\includegraphics[width=1\textwidth]{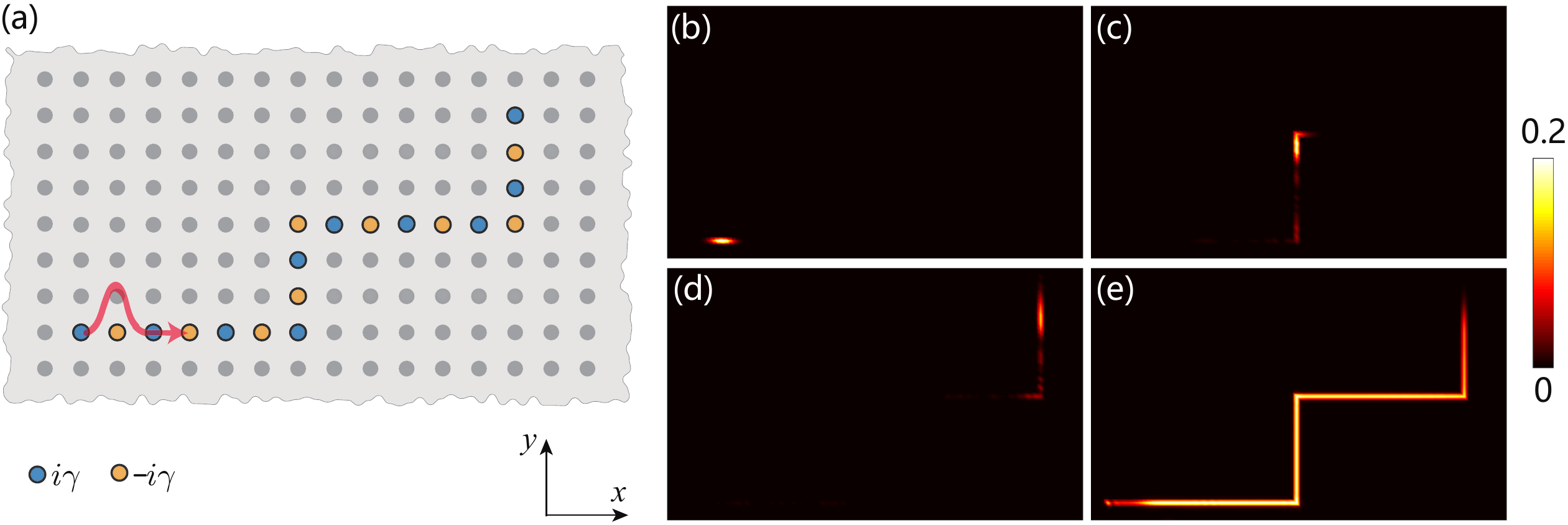}
\caption{(color online) (a) Schematic illustration of the waveguide path,
which is obtained by taking the nonzero non-Hermitian imaginary potentials $%
\pm i\protect\gamma $ along the path. (b)-(e) Numerical simulations of the
dynamics in waveguide. The initial state is a bilayer Gaussian wave packet.
Probability distribution at $t=0$, $t=T_{t}/2$ and $t=T_{t}$ are showed in
(b), (c) and (d), respectively. (e) The trace of the wave packet. Parameters
for the system are $t=1$, $\Delta =15$, and $\protect\gamma =11$; the size
of the system is $100\times 60\times 2$. Parameters for the initial state
Eq. (\protect\ref{initial_state}) are $\protect\alpha =0.4$, $N_{\mathrm{c}%
}=10$ and $k_{\mathrm{c}}=-\protect\pi /2$. The total duration of the
simulation is $T_{t}=42J^{-1}$, where $J$ is the scale of the Hamiltonian
and we take $J=1$.}
\label{Fig4}
\end{figure*}

\begin{figure*}[tbh]
\centering
\includegraphics[width=1\textwidth]{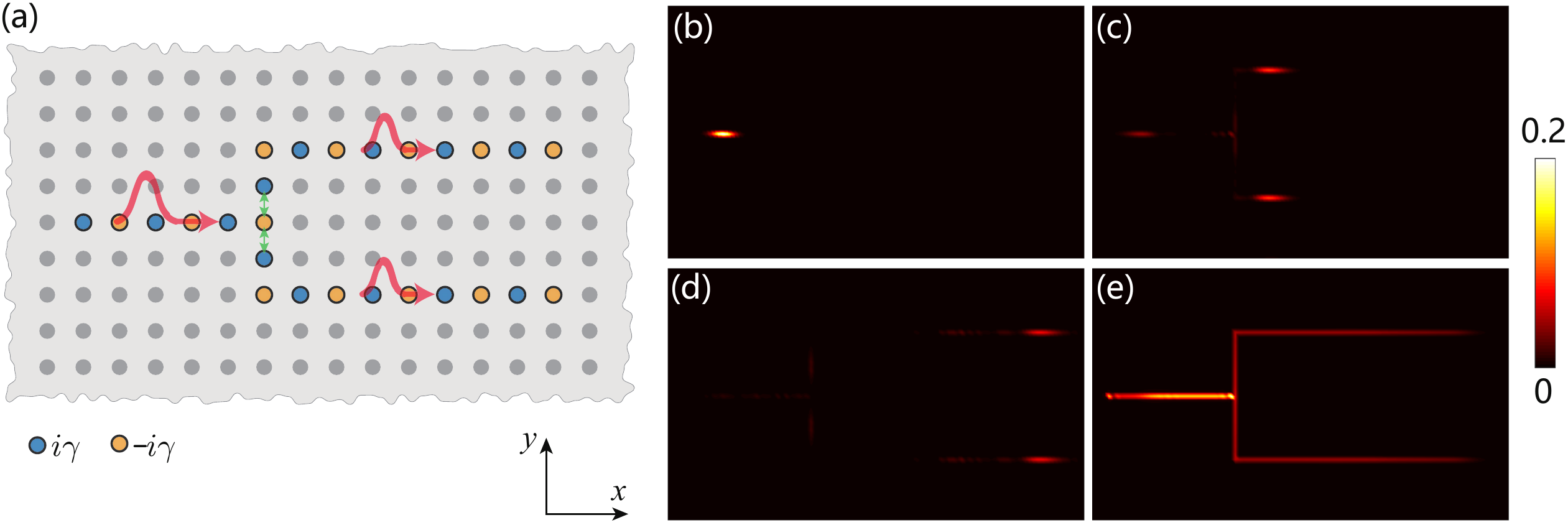}
\caption{(color online) (a) Schematic illustration of the beam splitter
waveguide. In order to get high transmission rate, we take $\protect\kappa %
\rightarrow \protect\kappa /\protect\sqrt{2}$ for the hopping (labeled by
green arrows) connecting the joint. (b)-(e) Numerical simulations of the
dynamics in waveguide. The initial state is a bilayer Gaussian wave packet.
Probability distribution at $t=0$, $t=T_{t}/2$ and $t=T_{t}$ are showed in
(b), (c) and (d), respectively. (e) The trace of the wave packet. Parameters
for the system are $t=1$, $\Delta =15$, and $\protect\gamma =11$; the size
of the system is $100\times 60\times 2$. Parameters for the initial state
Eq. ( \protect\ref{initial_state}) are $\protect\alpha =0.4$, $N_{\mathrm{c}%
}=10$ and $k_{\mathrm{c}}=-\protect\pi /2$. The total duration of the
simulation is $T_{t}=32J^{-1}$, where $J$ is the scale of the Hamiltonian
and we take $J=1$.}
\label{Fig5}
\end{figure*}

\section{Dynamics in waveguide}

\label{Dynamics in waveguide}

Before proceeding, we present a general non-Hermitian model which exhibits
Dirac-probability preserving dynamics. Considering a chiral symmetric
system, the Hamiltonian can be written in the block off-diagonal form \cite%
{Ryu}%
\begin{equation}
H=\left( 
\begin{array}{cc}
0 & D \\ 
D^{\dag } & 0%
\end{array}%
\right) ,  \label{hq}
\end{equation}%
where $D$\ as an arbitrary $N\times N$\ matrix. The basis in $H$ can be a
complete set of site states for a bipartite lattice. Based on $H$ as the
original Hermitian Hamiltonian, a non-Hermitian Hamiltonian $\mathcal{H}$ is
generated as the form%
\begin{equation}
\mathcal{H}=H+i\gamma \sigma _{z}\otimes I_{N},  \label{NH h}
\end{equation}%
where $\sigma _{z}$ is the Pauli matrix, and $I_{N}$ denotes the $N\times N$
identity matrix. The chiral symmetry of $H$ ensures its eigenvalues and
eigenvectors have the following properties: (i) The eigenvalues are always
in pairs, i.e., the spectrum has the form $\left\{ \varepsilon
_{n},\varepsilon _{-n}\right\} $ with $\varepsilon _{-n}=-\varepsilon _{n}$.
(ii) The corresponding eigenvector $\left\{ \left\vert \phi
_{n}\right\rangle ,\left\vert \phi _{-n}\right\rangle \right\} $ obeys 
\begin{equation}
\left\vert \phi _{n}\right\rangle =\left( \sigma _{z}\otimes I_{N}\right)
\left\vert \phi _{-n}\right\rangle .
\end{equation}

It turns out that (see Appendix 2) the spectrum of $\mathcal{H}$\ has the
form $\left\{ \epsilon _{n},\epsilon _{-n}\right\} $ with 
\begin{equation}
\epsilon _{\pm n}=\pm \left( \varepsilon _{n}^{2}-\gamma ^{2}\right) ^{1/2}
\end{equation}%
and the eigenvector of $\left\vert \varphi _{\pm n}\right\rangle $\ can be
mapped directly from $\left\vert \phi _{\pm n}\right\rangle $%
\begin{equation}
\left\vert \varphi _{\pm n}\right\rangle =M_{\pm n}\left\vert \phi _{\pm
n}\right\rangle  \label{State}
\end{equation}%
with the mapping matrix 
\begin{equation}
M_{\pm n}=\left( 
\begin{array}{cc}
a_{\pm n}I_{N} & 0 \\ 
0 & I_{N}%
\end{array}%
\right) ,  \label{Mapping}
\end{equation}%
where $a_{\pm n}=[\epsilon _{\pm n}+i\gamma ]/\varepsilon _{\pm n}$ is a
complex number with unit modulus, $\left\vert a_{\pm n}\right\vert =1$ for
real $\epsilon _{\pm n}$.

Obviously, $M_{\pm n}$\ is unitary matrix when the spectrum is fully real.
Remarkably, it can be proved that eigenvector set $\left\{ \left\vert
\varphi _{\pm n}\right\rangle \right\} $ obeys quasi-orthonormal relation
under the Dirac inner product, i.e., 
\begin{equation}
\langle \varphi _{m}\left\vert \varphi _{n}\right\rangle =\delta _{mn},
\end{equation}%
for $mn>0$ and $\epsilon _{m}\epsilon _{n}\neq 0$. It indicates that the
non-Hermitian system acts as a Hermitian one when only one of subspaces with
positive or negative spectrum is concerned. A direct conclusion is that, the
evolved state in one of the subspace of $\mathcal{H}$ maintains the
preservation of Dirac probability, exhibiting the Hermitian dynamic behavior.

Based on the above analysis, the bilayer system is a candidate for waveguide
allowing quantum transport with probability preserving. We take the
waveguide array by a collection of sites $\left\{ \left( j,l\right) \right\} 
$ and setting $\gamma _{j,l}=\gamma $, but zero for the rest of sites. In
this paper, the set of sites $\left\{ \left( j,l\right) \right\} $\ is
selected under the rule: it forms a non-Hermitian ladder system, described
by the Hamiltonian%
\begin{eqnarray}
H_{\mathrm{Ladd}} &=&\kappa \sum_{\lambda =1,2}\sum_{n=1}^{N}\alpha
_{n,\lambda }^{\dag }\alpha _{n+1,\lambda }+\Delta \sum_{n=1}^{N}\alpha
_{n,1}^{\dag }\alpha _{n,2}+\mathrm{H.c.}  \notag \\
&&+\sum_{n=1}^{N}\left( -1\right) ^{n}i\gamma (\alpha _{n,1}^{\dag }\alpha
_{n,1}-\alpha _{n,2}^{\dag }\alpha _{n,2}).  \label{ladder}
\end{eqnarray}%
The scheme is schematically illustrated in Fig. \ref{Fig3}(b). There are
numerous configurations for the path to connect any two distant locations.
When the parameters are in the range $\gamma \sim \Delta \gg \left\vert
\kappa \right\vert $, the sub-system $H_{\mathrm{Ladd}}$\ decouples from the
the bilayer system. The dynamics in the waveguide array is governed by the
Hamiltonian $H_{\mathrm{Ladd}}$. In certain parameter region, $H_{\mathrm{%
Ladd}}$\ possesses fully real spectrum and obeys the Hermitian dynamics
since it has the form of Eq. (\ref{NH h}).

The system $H_{\mathrm{Ladd}}$\ can be regarded as an extended non-Hermitian
SSH chain with long-range hopping term (see Appendix 3). It turns out that
such a model share the same dynamic behaviors with the simplest
non-Hermitian SSH chain within certain parameter region (strong dimerization
limit), which has been studied in the previous works \cite{HWH,ZKL1,ZKL2}.
It is expected that the dynamics of a non-Hermtian SSH chain emerges in the
bilayer squrare lattice with pre-engineered imaginary impurities.

To see waveguide dynamics in the bilayer square lattice system, we perform
the numerical simulations. The initial state is taken as the bilayer
Gaussian wave packet, which has the form 
\begin{equation}
\left\vert \psi \left( 0\right) \right\rangle =\Omega ^{-1/2}\sum_{\lambda
=1,2}\sum_{n=1}^{N}e^{-\alpha ^{2}\left( n-N_{\mathrm{c}}\right)
^{2}/2}e^{ik_{\mathrm{c}}n}\alpha _{n,\lambda }^{\dag }\left\vert \mathrm{vac%
}\right\rangle ,  \label{initial_state}
\end{equation}%
where $n$ is the site index along the waveguide array, $N_{\mathrm{c}}$ is
the center of Gaussian wave packet, $k_{\mathrm{c}}$ is the central
momentum, and $\Omega =2\sqrt{\pi }/\alpha $ is the Dirac normalization
factor. According to previous work \cite{HWH,ZKL1,ZKL2},\ the wave packet
should propagte along the array without spreading approximately, if the
sub-bands are sufficiently separated from the bilayer bands.\ The evolved
state has the form 
\begin{equation}
\left\vert \psi \left( t\right) \right\rangle =e^{-iHt}\left\vert \psi
\left( 0\right) \right\rangle ,
\end{equation}%
which can be computed by exact diagonalization.

The probability distribution at position $(i,j)$ at time $t$ is defined by
the sum of the Dirac probabilities of the top and bottom layers%
\begin{equation}
p(i,j,t)=\sum_{\lambda =1,2}\left\vert \langle i,j,\lambda \left\vert \psi
\left( t\right) \right\rangle \right\vert ^{2}.  \label{p_ijt}
\end{equation}%
where $\left\vert i,j,\lambda \right\rangle $ denotes the position state. To
demonstrate the efficiency of the waveguide, we define the function 
\begin{equation}
P(i,j)=\frac{p(i,j,t)_{\max }}{\left[ \sum_{t}p(i,j,t)\right] _{\max }}%
\sum_{t}p(i,j,t),  \label{trace}
\end{equation}%
to record the trace of the wave packet. The simulation is performed for the
systems far from EP, i.e., the positive and negative waveguide bands are
well separated, and two kinds of waveguide configurations: (i) a single path
waveguide with two right-angle bends, and\ (ii) a beam splitter with two
right-angle bends. Their schematics are shown in Fig. \ref{Fig4}(a) and Fig. %
\ref{Fig5}(a). The initial state is a Gaussian wave packet with central
momentum $k_{\mathrm{c}}=-\pi /2$. Since it only relates to a single sub-band in large gap limit,
it is expected that the time evolution exhibits a probability preserving behavior approximately.
For case (i), the numerical simulation results indicates that the wave
packet propagate along the designed waveguide array path as expected. For
the (ii), it shows that the wave packet split into two parts after passing
through the joint. In both cases, the wave packet is confined within the
defects well and transmitted efficiently around the corners. These numerical
results demonstrate and verify our theory for quantum transport in
non-Hermitian impurity array.

\section{Conclusion and Discussion}

\label{Conclusion and Discussion}

In summary, we have studied the formation of band gap bound states induced
by a non-Hermitian impurity embedded in a Hermitian system. We have shown
that a pair of bound states emerges inside the band gap when a $\mathcal{PT}$
imaginary potential is added in a strongly coupled bilayer and the bound
states become strongly localized when the system approaches to the
exceptional point. Inspired by this, we construct an impurity array which
can be described by a non-Hermitian SSH type effective Hamiltonian which
possesses $\mathcal{CT}$\ symmetry. It is the first time to establish a
theory for non-Hermitian impurity-induced waveguide, which paves the way for
the non-Hermitian device design. As a supplementary material (Appendix 2)
for our theory, we first provide a rigorous proof for the features of a $%
\mathcal{CT}$-symmetric system, including the reality spectrum and quasi
orthogonality of Dirac inner product. We would like to point out that such
two conclusions are independent of other requirements for the original Hermitian system,
such as translational symmetry.  Thus a variety of non-Hermitian models with various of geometries is allowed. The numerical simulations for the quantum transport of wave
packet in right-angle bends waveguide and $Y$-beam splitter have
demonstrated this point.

\section*{Appendix}

\setcounter{equation}{0} \renewcommand{\theequation}{A\arabic{equation}} %
\renewcommand{\thesubsection}{\arabic{subsection}}

In this appendix we present the Bethe ansatz solution for the Hamiltonian
from Eq. (\ref{Chain_H}) and properties of the non-Hermitian ladder system.

\subsection{Bound states induced by non-Hermitian impurity}

For a single-tunneling non-Hermitian Hamiltonian (\ref{Chain_H}), the
bound-state Bethe ansatz wave function has the form 
\begin{equation}
\left\vert \psi _{\mathrm{B}}\right\rangle =\sum_{\sigma =\pm
}\sum_{j=-\infty }^{\infty }C^{\sigma }\left( -\sigma \right) ^{\left\vert
j\right\vert }e^{-\beta ^{\sigma }\left\vert j\right\vert }a_{j,\sigma
}^{\dag }\left\vert \mathrm{vac}\right\rangle ,
\end{equation}%
where $\beta ^{\sigma }$ is a positive real number, indicating the strength
of localization around the non-Hermitian tunneling $iT$. The Schr\"{o}dinger
equation $H_{\mathrm{Chain}}\left\vert \psi _{\mathrm{B}}\right\rangle =E_{%
\mathrm{B}}\left\vert \psi _{\mathrm{B}}\right\rangle $ with bound-state
energy $E_{\mathrm{B}}$\ gives 
\begin{eqnarray}
E_{\mathrm{B}} &=&-2\kappa \cosh \beta ^{+}+\Delta  \notag \\
&=&2\kappa \cosh \beta ^{-}-\Delta ,  \label{E_B1}
\end{eqnarray}%
at $\left\vert j\right\vert \geqslant 1$, and 
\begin{equation}
\left\{ 
\begin{array}{c}
\left( \Delta -2e^{-\beta ^{+}}\kappa -E_{\mathrm{B}}\right) C^{+}+iTC^{-}=0
\\ 
iTC^{+}+\left( -\Delta +2e^{-\beta ^{-}}\kappa -E_{\mathrm{B}}\right) C^{-}=0%
\end{array}%
\right. ,  \label{Cpm}
\end{equation}%
at $j=0$. The existence of bound-state solution requires 
\begin{equation}
\left\vert 
\begin{array}{cc}
\left( \Delta -2e^{-\beta ^{+}}\kappa -E_{\mathrm{B}}\right) & iT \\ 
iT & \left( -\Delta +2e^{-\beta ^{-}}\kappa -E_{\mathrm{B}}\right)%
\end{array}%
\right\vert =0.  \label{Det}
\end{equation}%
Then evanescent coefficient $\beta ^{\sigma }$\ can be determined by 
\begin{equation}
\left\{ 
\begin{array}{c}
\cosh \beta ^{-}+\cosh \beta ^{+}=\Delta /\kappa \\ 
\sinh \beta ^{+}\sinh \beta ^{-}=T^{2}/(2\kappa )^{2}%
\end{array}%
\right. ,  \label{eq_beta}
\end{equation}%
And Eq. (\ref{Cpm}) leads to 
\begin{equation}
C^{\sigma }=\frac{2\sigma \kappa }{iT}\sinh \beta ^{-\sigma }C^{-\sigma }.
\end{equation}%
Then the wave function can be written as 
\begin{eqnarray}
\left\vert \psi _{\mathrm{B}}\right\rangle &=&\sum_{j=-\infty }^{\infty
}\left( e^{i\pi j}e^{-\beta ^{+}\left\vert j\right\vert }a_{j,+}^{\dag
}\right) \left\vert \mathrm{vac}\right\rangle  \notag \\
&&-\frac{2\kappa }{iT}\sinh \beta ^{+}\sum_{j=-\infty }^{\infty }\left(
e^{-\beta ^{-}\left\vert j\right\vert }a_{j,-}^{\dag }\right) \left\vert 
\mathrm{vac}\right\rangle ,
\end{eqnarray}%
where the normalization coefficient is neglected since it should be valued
in the frameworks of Dirac or biothonormal inner product\ in practice. We
note that Eq. (\ref{eq_beta}) is symmetric under the operation $\beta
^{+}\leftrightarrows \beta ^{-}$. Thus, if $(\beta ^{+},\beta ^{-})=(x,y)$
is a solution with the eigenvalue $E_{\mathrm{B}}$, then $(\beta ^{+},\beta
^{-})=(y,x)\ $corresponds to another solution with eigenvalue $-E_{\mathrm{B}%
} $. Accordingly, when take $T=T_{\mathrm{c}}=\sqrt{\Delta ^{2}-(2\kappa
)^{2}} $, we have $\beta _{c}=\beta ^{+}=\beta ^{-}$ with 
\begin{equation}
\beta _{c}=\ln \left[ \Delta /\left( 2\kappa \right) +\sqrt{\Delta
^{2}/\left( 4\kappa ^{2}\right) -1}\right]
\end{equation}%
and two bound states coalesce to a single state 
\begin{equation}
\left\vert \psi _{\mathrm{B}}^{\mathrm{EP}}\right\rangle =\sum_{j=-\infty
}^{\infty }\left( e^{i\pi j}e^{-\beta _{c}\left\vert j\right\vert
}a_{j,+}^{\dag }+ie^{-\beta _{c}\left\vert j\right\vert }a_{j,-}^{\dag
}\right) \left\vert \mathrm{vac}\right\rangle ,  \label{psi_EP}
\end{equation}%
with the eigenenergy $E_{\mathrm{B}}^{\mathrm{EP}}=0$, indicating the
occurrence of EP. We note that $\left\vert \psi _{\mathrm{B}}^{\mathrm{EP}%
}\right\rangle $\ has the identical Dirac probability distributions on the
two chains.

In strong localization limit with $e^{2\beta ^{\sigma }}\gg 1$, we have the
approximate solutions%
\begin{eqnarray}
\left( 
\begin{array}{c}
\beta ^{-} \\ 
\beta ^{+}%
\end{array}%
\right) &=&\left( 
\begin{array}{c}
\ln (\Delta /\kappa +\sqrt{\Delta ^{2}-T^{2}}/\kappa ) \\ 
\ln (\Delta /\kappa -\sqrt{\Delta ^{2}-T^{2}}/\kappa )%
\end{array}%
\right) ,  \notag \\
&&\text{or }\left( 
\begin{array}{c}
\ln (\Delta /\kappa -\sqrt{\Delta ^{2}-T^{2}}/\kappa ) \\ 
\ln (\Delta /\kappa +\sqrt{\Delta ^{2}-T^{2}}/\kappa )%
\end{array}%
\right) ,
\end{eqnarray}%
which are still in agreement with the symmetry of Eq. (\ref{eq_beta}).

\subsection{Dirac probability preservation}

Consider a Hermitian Hamiltonian with chiral symmetry,%
\begin{equation}
H=\left( 
\begin{array}{cc}
0 & D \\ 
D^{\dagger } & 0%
\end{array}%
\right) ,
\end{equation}%
where $D$ as an arbitrary $N\times N$ matrix. The Schr\"{o}dinger equation
is 
\begin{equation}
H\left\vert \phi _{\pm n}\right\rangle =\varepsilon _{\pm n}\left\vert \phi
_{\pm n}\right\rangle
\end{equation}%
and obey%
\begin{equation}
\varepsilon _{-n}=-\varepsilon _{n},\left\vert \phi _{n}\right\rangle
=\left( \sigma _{z}\otimes I_{N}\right) \left\vert \phi _{-n}\right\rangle ,
\end{equation}%
due to the chiral symmetry%
\begin{equation}
\left\{ \sigma _{z}\otimes I_{N},H\right\} =0.
\end{equation}%
A non-Hermitian Hamiltonian $\mathcal{H}$ can be generated as the form%
\begin{equation}
\mathcal{H}=H+i\gamma \sigma _{z}\otimes I_{N},
\end{equation}%
where $\sigma _{z}$ is the Pauli matrix, and $I_{N}$ denotes the $N\times N$
identity matrix. We note that non-Hermitian term breaks the chiral symmetry,
but $\mathcal{H}$\ has $\mathcal{CT}$ symmetry.\ The corresponding Schr\"{o}%
dinger equation is 
\begin{equation}
\mathcal{H}\left\vert \varphi _{\pm n}\right\rangle =\epsilon _{\pm
n}\left\vert \varphi _{\pm n}\right\rangle ,
\end{equation}%
and obey 
\begin{equation}
\epsilon _{\pm n}=\pm \left( \varepsilon _{n}^{2}-\gamma ^{2}\right)
^{1/2},\left\vert \varphi _{\pm n}\right\rangle =M_{\pm n}\left\vert \phi
_{\pm n}\right\rangle ,
\end{equation}%
with the mapping matrix 
\begin{equation}
M_{\pm n}=\left( 
\begin{array}{cc}
a_{\pm n}I_{N} & 0 \\ 
0 & I_{N}%
\end{array}%
\right) ,
\end{equation}%
where $a_{\pm n}=[\epsilon _{\pm n}+i\gamma ]/\varepsilon _{\pm n}$ fulfills 
$\left\vert a_{\pm n}\right\vert =1$ for real $\epsilon _{\pm n}$ and is
pure imaginary for imaginary $\epsilon _{\pm n}$. For real $\epsilon _{\pm
n} $, the factor $a_{\pm n}$ can be written in the form of $a_{\pm
n}=e^{i\theta _{n}}$, with $\theta _{n}=\arctan \left( \gamma /\epsilon
_{\pm n}\right) $. This can be shown as following.

Actually, from%
\begin{equation}
HM_{\pm n}=\left( 
\begin{array}{cc}
I_{N} & 0 \\ 
0 & a_{\pm n}I_{N}%
\end{array}%
\right) H,
\end{equation}%
we have $H\left\vert \varphi _{\pm n}\right\rangle =\left( HM_{\pm n}\right)
\left\vert \phi _{\pm n}\right\rangle $, and%
\begin{eqnarray}
H\left\vert \varphi _{\pm n}\right\rangle &=&\varepsilon _{\pm n}\left( 
\begin{array}{cc}
I_{N} & 0 \\ 
0 & a_{\pm n}I_{N}%
\end{array}%
\right) \left\vert \phi _{\pm n}\right\rangle \\
&=&\varepsilon _{\pm n}\left( 
\begin{array}{cc}
\left( a_{\pm n}\right) ^{-1}I_{N} & 0 \\ 
0 & a_{\pm n}I_{N}%
\end{array}%
\right) \left\vert \varphi _{\pm n}\right\rangle .  \notag
\end{eqnarray}%
Therefore, from%
\begin{equation}
\mathcal{H}\left\vert \varphi _{\pm n}\right\rangle =\left( H+i\gamma \sigma
_{z}\otimes I_{N}\right) \left\vert \varphi _{\pm n}\right\rangle ,
\end{equation}%
and $a_{\pm n}=\left( \epsilon _{\pm n}+i\gamma \right) /\varepsilon _{\pm
n} $, $\epsilon _{\pm n}=\pm \left( \varepsilon _{n}^{2}-\gamma ^{2}\right)
^{1/2}$, we have $\varepsilon _{\pm n}\left( a_{\pm n}\right) ^{-1}+i\gamma
=\varepsilon _{\pm n}a_{\pm n}-i\gamma =\epsilon _{\pm n}$, which leads to%
\begin{equation}
\mathcal{H}\left\vert \varphi _{\pm n}\right\rangle =\epsilon _{\pm
n}\left\vert \varphi _{\pm n}\right\rangle .
\end{equation}%
The mapping relation between vectors $\left\vert \phi _{n}\right\rangle $\
and $\left\vert \varphi _{n}\right\rangle $ can result in many interesting
and useful applications. In general, set $\left\{ \left\vert \phi
_{n}\right\rangle \right\} $ obeys orthogonal relation in the framework of
Dirac inner product, while set $\left\{ \left\vert \varphi _{n}\right\rangle
\right\} $\ does not due to the non-Hermiticity of $\mathcal{H}$. We will show
that $\left\{ \left\vert \varphi _{n}\right\rangle \right\} $\ still obeys
Dirac orthogonal relation within one of the subspace ($n>0$ or $n<0$).

We start with the eigenvector of Hermitian Hamiltonian $H$, which has the
form 
\begin{equation}
\left\vert \phi _{n}\right\rangle =\left( 
\begin{array}{c}
A_{n} \\ 
B_{n}%
\end{array}%
\right) ,
\end{equation}%
where $A_{n}$ and $B_{n}$ are two $N\times 1$ vectors, representing the wave
function of sublattice $\mathrm{A}$ and $\mathrm{B}$. $\left\{ \left\vert
\phi _{n}\right\rangle \right\} $ fulfill orthogonal normalization condition 
\begin{equation}
\left\langle \phi _{m}\right. \left\vert \phi _{n}\right\rangle =A_{m}^{\dag
}A_{n}+B_{m}^{\dag }B_{n}=\delta _{mn}.  \label{norm}
\end{equation}%
The Schr\"{o}dinger equation of $\left\vert \phi _{n}\right\rangle $ has the
form 
\begin{equation}
\left( 
\begin{array}{cc}
0 & D \\ 
D^{\dagger } & 0%
\end{array}%
\right) \left( 
\begin{array}{c}
A_{n} \\ 
B_{n}%
\end{array}%
\right) =\varepsilon _{n}\left( 
\begin{array}{c}
A_{n} \\ 
B_{n}%
\end{array}%
\right) ,
\end{equation}%
or explicitly 
\begin{equation}
\left\{ 
\begin{array}{c}
DB_{n}=\varepsilon _{n}A_{n} \\ 
D^{\dagger }A_{n}=\varepsilon _{n}B_{n}%
\end{array}%
\right. .
\label{A_B_Eq}
\end{equation}%
Multiplying by $A_{m}^{\dag }$ or $B_{m}^{\dag }$, respectively, we have 
\begin{equation}
\left\{ 
\begin{array}{c}
A_{m}^{\dag }DB_{n}=\varepsilon _{n}A_{m}^{\dag }A_{n} \\ 
B_{m}^{\dag }D^{\dagger }A_{n}=\varepsilon _{n}B_{m}^{\dag }B_{n}%
\end{array}%
\right. .  \label{ADB}
\end{equation}%
Considering the conjugation of above Schr\"{o}dinger equation (\ref{A_B_Eq})
\begin{equation}
\left\{ 
\begin{array}{c}
B_{m}^{\dag }D^{\dag }=\varepsilon _{m}A_{m}^{\dag } \\ 
A_{m}^{\dag }D=\varepsilon _{m}B_{m}^{\dag }%
\end{array}%
\right. ,
\label{A_B_Eq_dagger}
\end{equation}%
Eq. (\ref{ADB}) together with Eq. (\ref{A_B_Eq_dagger}) gives 
\begin{equation}
\left\{ 
\begin{array}{c}
\varepsilon _{m}B_{m}^{\dag }B_{n}=\varepsilon _{n}A_{m}^{\dag }A_{n} \\ 
\varepsilon _{m}A_{m}^{\dag }A_{n}=\varepsilon _{n}B_{m}^{\dag }B_{n}%
\end{array}%
\right. ,
\end{equation}%
and consequently leads to 
\begin{equation}
\left( \varepsilon _{n}+\varepsilon _{m}\right) \left( A_{m}^{\dag
}A_{n}-B_{m}^{\dag }B_{n}\right) =0.
\end{equation}%
In the case of $mn>0$\ and $\varepsilon _{n}\varepsilon _{m}\neq 0$,\ we
have 
\begin{equation}
A_{m}^{\dag }A_{n}-B_{m}^{\dag }B_{n}=0.
\end{equation}%
Together with the orthogonal normalization condition Eq. (\ref{norm}), we
obtain 
\begin{equation}
A_{m}^{\dag }A_{n}=B_{m}^{\dag }B_{n}=\frac{1}{2}\delta _{mn},
\end{equation}%
which means that eigenvector $\left\vert \phi _{n}\right\rangle $ has the
same Dirac probability in sublattice $\mathrm{A}$ and $\mathrm{B}$.

This factor leads to an important conclusion for vector $\left\{ \left\vert
\varphi _{n}\right\rangle \right\} $: different eigenstates in one of the
subspace ($mn>0$) of $\mathcal{H}$ are still orthogonal. In fact, the
mapping matrix gives 
\begin{eqnarray}
\langle \varphi _{m}\left\vert \varphi _{n}\right\rangle &=&\left\langle
\phi _{m}\right\vert M_{m}^{\dag }M_{n}\left\vert \phi _{n}\right\rangle 
\notag \\
&=&a_{m}^{\ast }a_{n}A_{m}^{\dag }A_{n}+B_{m}^{\dag }B_{n}  \notag \\
&=&\frac{1}{2}\left( a_{m}^{\ast }a_{n}+1\right) \delta _{mn}.
\end{eqnarray}
In addition, for full real spectrum we have 
\begin{equation}
\langle \varphi _{m}\left\vert \varphi _{n}\right\rangle =\delta _{mn},
\end{equation}%
since $M_{n}$ is unitary, i.e., $\left\vert a_{n}\right\vert =$ $\left\vert
e^{i\theta _{n}}\right\vert =1$.

As an application, we will show that for any initial state in one of the
subspace ($n>0$ or $n<0$) of the non-Hermitian system $\mathcal{H},$ the
time evolution preserves the Dirac probability. Considering an initial state
in one of the subspace with the form $\left\vert \psi (t=0)\right\rangle
=\sum_{n=1}^{N}C_{n}\left\vert \varphi _{n}\right\rangle $ with $%
\sum_{n=1}^{N}C_{n}^{\ast }C_{n}=1$, the evolved state can be written as 
\begin{equation}
\left\vert \psi \left( t\right) \right\rangle =\sum_{n=1}^{N}C_{n}\exp
\left( -i\epsilon _{n}t\right) \left\vert \varphi _{n}\right\rangle .
\end{equation}%
In the condition of full real spectrum, the Dirac probability is 
\begin{eqnarray}
P\left( t\right) &=&\langle \psi \left( t\right) \left\vert \psi \left(
t\right) \right\rangle  \notag \\
&=&\sum_{m,n=1}^{N}C_{m}^{\ast }C_{n}\exp \left[ i\left( \epsilon
_{m}-\epsilon _{n}\right) t\right] \left\langle \varphi _{m}\right.
\left\vert \varphi _{n}\right\rangle  \notag \\
&=&1,
\end{eqnarray}%
which maintains the preservation of Dirac probability, exhibiting the
Hermitian dynamic behavior.

\subsection{Non-Hermitian ladder and SSH models}

In the condition of $\Delta \gg \kappa $, (strong dimerization limit) the
non-Hermitian ladder described by the Hamiltonian in Eq. (\ref{ladder}) 
\begin{eqnarray}
H_{\mathrm{Ladd}} &=&\kappa \sum_{\lambda =1,2}\sum_{j=1}^{N}\alpha
_{j,\lambda }^{\dag }\alpha _{j+1,\lambda }+\Delta \sum_{j=1}^{N}\alpha
_{j,1}^{\dag }\alpha _{j,2}+\mathrm{H.c.}  \notag \\
&&+\sum_{j}\left( -1\right) ^{n}i\gamma (\alpha _{j,1}^{\dag }\alpha
_{j,1}-\alpha _{j,2}^{\dag }\alpha _{j,2}),
\end{eqnarray}%
is equivalent to a non-Hermitian SSH model%
\begin{eqnarray}
H_{\mathrm{SSH}} &=&\sum_{j=1}^{N}(\Delta a_{j}^{\dag }b_{j}+2\kappa
a_{j+1}^{\dag }b_{j})+\mathrm{H.c.}  \notag \\
&&+i\gamma \sum_{j=1}^{N}\left( a_{l}^{\dag }a_{l}-b_{l}^{\dag }b_{l}\right)
.
\end{eqnarray}%
Actually, the core matrix of $H_{\mathrm{Ladd}}$\ is 
\begin{equation}
h_{k}^{\mathrm{Ladd}}=\left( 
\begin{array}{cc}
i\gamma  & \Delta +2\kappa \cos k \\ 
\Delta +2\kappa \cos k & -i\gamma 
\end{array}%
\right) .  \label{ladder_h}
\end{equation}%
\begin{figure}[tbp]
\centering
\includegraphics[width=0.48\textwidth]{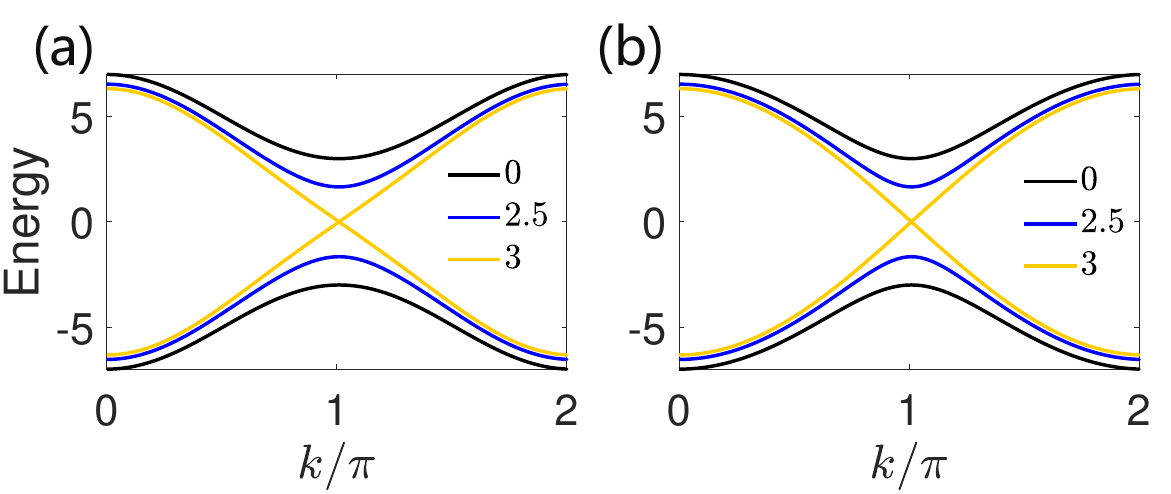}
\caption{(color online) Spectra of (a) ladder model and (b) SSH model with
the core matrix Eq. (\protect\ref{ladder_h}) and Eq. (\protect\ref{ssh_h})
for different $\protect\gamma $. The Parameters are $\protect\kappa =1$, $%
\Delta =5$, and $\protect\gamma =0,2.5,3$. }
\label{Fig6}
\end{figure}

For $H_{\mathrm{SSH}}$, we have the core matrix%
\begin{equation}
h_{k}^{\mathrm{SSH}}=\left( 
\begin{array}{cc}
i\gamma & \Delta +2\kappa e^{-ik} \\ 
\Delta +2\kappa e^{ik} & -i\gamma%
\end{array}%
\right) ,  \label{ssh_h}
\end{equation}%
which has the same spectrum with 
\begin{equation}
h_{k}=\left( 
\begin{array}{cc}
i\gamma & \sqrt{\Delta ^{2}+4\kappa ^{2}+4\Delta \kappa \cos k} \\ 
\sqrt{\Delta ^{2}+4\kappa ^{2}+4\Delta \kappa \cos k} & -i\gamma%
\end{array}%
\right) .
\end{equation}%
In the case of strong dimerization limit $\Delta ^{2}\gg \kappa ^{2}$, we
have%
\begin{equation}
\sqrt{\Delta ^{2}+4\kappa ^{2}+4\Delta \kappa \cos k}\approx \Delta +2\kappa
\cos k,
\end{equation}%
or%
\begin{equation}
h_{k}\approx h_{k}^{\mathrm{Ladd}}.
\end{equation}

The spectra of ladder model and SSH model with several $\gamma$ are shown in
Fig. \ref{Fig6}(a) and (b), respectively. It indicate that two spectra are
almost identical.

\acknowledgments This work was supported by National Natural Science
Foundation of China (under Grant No. 11874225).

\end{document}